\documentclass[twocolumn]{aastex62}

\usepackage{natbib}
\graphicspath{{./}{figures/}}
\usepackage{graphicx}
\usepackage{float}

\begin{document}

\title{The Age Evolution of the Radio Morphology of Supernova Remnants}

\correspondingauthor{Jennifer N. Stafford}
\email{stafford.205@osu.edu}

\author{Jennifer N. Stafford}
\affiliation{Department of Astronomy, The Ohio State University, 140 W. 18th Ave., Columbus, Ohio 43210, USA}

\author{Laura A. Lopez}
\affiliation{Department of Astronomy, The Ohio State University, 140 W. 18th Ave., Columbus, Ohio 43210, USA}
\affil{Center for Cosmology and AstroParticle Physics, The Ohio State University, 191 W. Woodruff Ave., Columbus, OH 43210, USA}
\affil{Niels Bohr Institute, University of Copenhagen, Blegdamsvej 17, 2100 Copenhagen, Denmark}

\author{Katie Auchettl}
\affil{Center for Cosmology and AstroParticle Physics, The Ohio State University, 191 W. Woodruff Ave., Columbus, OH 43210, USA}
\affil{Department of Physics, The Ohio State University, 191 W. Woodruff Ave., Columbus, OH 43210, USA}

\author{Tyler Holland-Ashford}
\affiliation{Department of Astronomy, The Ohio State University, 140 W. 18th Ave., Columbus, Ohio 43210, USA}

\begin{abstract}

Recent hydrodynamical models of supernova remnants (SNRs) demonstrate that their evolution depends heavily on the inhomogeneities of the surrounding medium. As SNRs expand, their morphologies are influenced by the non-uniform and turbulent structure of their environments, as reflected in their radio continuum emission. In this paper, we measure the asymmetries of 96 SNRs in radio continuum images from three surveys of the Galactic plane and compare these results to the SNRs' radii, which we use as a proxy for their age. We find that larger (older) SNRs are more elliptical/elongated and more mirror asymmetric than smaller (younger) SNRs, though the latter vary in their degrees of asymmetry. This result suggests that SNR shells become more asymmetric as they sweep up the interstellar medium (ISM), as predicted in hydrodynamical models of SNRs expanding in a multi-phase or turbulent ISM. 

\end{abstract}


\section{Introduction} \label{sec:intro}

Synchrotron radiation is produced by electrons that are accelerated at the shocks of supernova remnants (SNRs) and interact with the local magnetic field (see e.g., \citealt{berezhko04}). This emission dominates at radio wavelengths, particularly at lower frequencies (e.g. 1.4~GHz) where thermal bremsstrahlung contributes less to the spectrum. Radio continuum surveys are the primary means by which new SNRs are identified (see e.g., \citealt{chomiuk09} and references therein). At present, 295 objects have been classified as SNRs in the Milky Way \citep{green17}, and 95\% of these are radio sources \citep{dubner15}. 

Historically, SNRs have been classified based on their non-thermal radio shells (e.g. \citealt{milne69}). After many SNRs were resolved at radio wavelengths, astronomers began to categorize them based on their morphologies (e.g., \citealt{green84}). The categories most commonly used are shell-type, composite, and mixed-morphology. Shell-type SNRs are those with limb-brightened radio shell emission (e.g., SN~1006: \citealt{winkler14}, G1.9$+$0.3: \citealt{dehorta14}) and 79\% of Galactic SNRs fall into this group \citep{dubner15}. Composite SNRs are ones that have both the shell and the center-filled emission from a pulsar wind nebula) (e.g., MSH 15-56: \citealt{dickel00}, Vela SNR: \citealt{dodson03}). Additionally, X-ray imaging led to another class of SNRs called mixed-morphology or thermal-composite \citep{rho98,lazendic06,shelton04} that have radio shells with center-filled X-rays (e.g., W44, IC443: \citealt{kawasaki05}). 

\begin{deluxetable*}{llcrcrrrc}[ht]
\footnotesize
\tablenum{1}
\tablecaption{List of SNRs in Our Sample from THOR\label{tab:tableTHOR}}
\tablehead{
\colhead{No.} & \colhead{Source\tablenotemark{a}} & \colhead{Alternate Names} & \colhead{Distance\tablenotemark{b}} & \colhead{Evidence of} & \colhead{$P_{1}/P_{0}$} & \colhead{$P_{2}/P_{0}$} & \colhead{$P_{3}/P_{0}$} & \colhead{References\tablenotemark{d}} \\
\colhead{} & \colhead{} & \colhead{} & \colhead{(kpc)}
& \colhead{Explosion Type\tablenotemark{c}} & \colhead{($\times10^{-5}$)} & \colhead{($\times10^{-7}$)} & \colhead{($\times10^{-7}$)} & \colhead{}}
\startdata
1 & G15.9$+$0.2 & & 8.5* & N & 84.9$_{-7.2}^{+9.0}$ & 8.50$_{-6.37}^{+6.70}$ & 4.90$_{-2.83}^{+2.46}$ & 1 \\ 
2 & G16.7$+$0.1$^\dagger$ & & 10 & N  & 3.01$\pm{1.3}$ & 18.7$_{-8.3}^{+7.3}$ & 2.79$_{-1.82}^{+1.84}$ &  2 \\
3 & G18.1$-$0.1$^\dagger$ & & 6.4$\pm{0.2}$  & E & 95.9$_{-1.4}^{+1.5}$ & 326$\pm{9}$ & 5.25$_{-0.64}^{+0.51}$ & 3 \\
4& G18.8$+$0.3$^\dagger$ & Kes~67 & 13.8$\pm{4.0}$ & E & 240$_{-2}^{+1}$ & 1119$_{-13}^{+12}$ & 149$\pm{2}$ & 3,4 \\
5&G20.0$-$0.2 & & 11.2$\pm{0.3}$ & E, N & 27.3$\pm{0.7}$ & 43.9$_{-2.8}^{+3.3}$  & 2.71$_{-0.47}^{+0.57}$ & 3 \\
6&G20.4$+$0.1 & & 7.8 & E, N & 74.9$\pm{2.9}$ & 259$_{-18}^{+19}$ & 33.4$_{-2.6}^{+2.4}$ & 5 \\
7&G21.5$-$0.1 & & 8.5* & -- & 13.1$_{-2.7}^{+2.2}$ & 16.2$_{-11.7}^{+12.5}$ & 2.74$_{-1.94}^{+1.62}$ & -- \\
8& G21.8$-$0.6$^\dagger$ & Kes~69 & 5.2$_{-0.3}^{+0.0}$ & E & 370$\pm{1}$ & 760$\pm{5}$ & 100$\pm{1}$ & 6, 7 \\
9&G22.7$-$0.2 & & 4.4$\pm{0.4}$ & E & 36.4$_{-0.2}^{+0.3}$ & 162$\pm{2}$ & 6.03$_{-0.20}^{+0.17}$ & 8\\
10&G23.3$-$0.3$^\dagger$ & W41 & 4.4$\pm{4}$ & E & 96.5$_{-0.3}^{+0.4}$ & 951$_{-4}^{+5}$ & 94.5$\pm{7}$ & 9, 10\\
11 &G27.4$+$0.0$^\dagger$ & Kes~73 & 5.8$\pm{0.3}$ &  N, A & 16.2$\pm{0.4}$ & 85.2$_{-3}^{+4}$ & 3.64$_{-0.32}^{+0.38}$ & 3, 11\\
12&G28.6$-$0.1 & & 9.6$\pm{0.3}$  & E & 3.82$_{-0.23}^{+0.21}$ & 819$\pm{12}$ & 189$\pm{3}$ & 5\\
13&G29.6$+$0.1$^\dagger$ & & 10.0$\pm{5.0}$ & N & 5.95$_{-1.12}^{+1.13}$ & 81.3$_{-15.9}^{+12.4}$ & 17.2$_{-3.4}^{+3.0}$ & 12 \\
14& G31.9$+$0.0$^\dagger$ & 3C~391 & 7.1$\pm{0.4}$  & E, A & 63.9$\pm{0.3}$ & 240$_{-2}^{+3}$ & 6.54$_{-0.18}^{+0.17}$ & 13, 14\\
15&G32.4$+$0.1$^\dagger$ & & 17 & E & 48.5$_{-5.4}^{+4.9}$ & 170$_{-34}^{+28}$ & 92.5$_{-14.2}^{+14.6}$ & 14, 15\\
16 & G32.8$-$0.1$^\dagger$ & Kes~78 & 4.8 & E, N & 87.9$_{-1.6}^{+1.3}$ & 4005$_{-50}^{+53}$ & 94.1$_{-2.5}^{+3.0}$ & 16, 17\\
17&G33.2$-$0.6$^\dagger$ & & 8.5* & -- & 180$_{-4}^{+3}$ & 106$\pm{8}$ & 0.34$_{-0.24}^{+0.25}$ & --\\
18&G33.6$+$0.1$^\dagger$ & Kes~79 & 3.5$\pm{0.3}$ & N, A  & 96.8$_{-0.9}^{+0.8}$ & 3.70$_{-0.50}^{+0.60}$ & 24.2$\pm{0.6}$ & 3,18,19\\
19 & G34.7$-$0.4$^\dagger$ & W44 & 3.0$\pm{0.3}$ & E, N, A  & 5.28$_{-0.03}^{+0.04}$ & 575$\pm{2}$ & 27.0$\pm{0.2}$ & 3,20,21 \\
20 & G35.6$-$0.4 &  & 3.6$\pm{0.4}$  &  E & 5.21$_{-0.20}^{+0.23}$ & 538$\pm{9}$ & 5.28$_{-0.53}^{+0.47}$ & 22\\
21 & G36.6$-$0.7 & & 8.5* & -- & 190$\pm{3}$ & 825$\pm{22}$ & 34.8$_{-2.8}^{+2.9}$ & -- \\
22 & G49.2$-$0.7$^\dagger$ & W51C & 5.4$\pm{0.6}$  & E, A & 8.56$_{-0.09}^{+0.08}$ & 726$_{-3}^{+2}$ & 50.7$\pm{0.3}$ & 23, 24\\
\enddata 
\tablenotetext{a}{ $\dagger$ Denotes SNRs with evidence of interaction with a molecular cloud: G16.7$+$0.1: \citealt{green97} (G97); \citealt{reynosomangum00}; \citealt{hewitt08} (H08); \citealt{kilpatrick16} (K16); G18.1$-$0.1: \citealt{froebrich15} (F15); G18.8$+$0.3: \citealt{dubner04,tian07}; G21.8$-$0.6: G97; H08; \citealt{zhou09,hewitt09b}; F15; G23.3$-$0.3: \citealt{frail13} G27.4$+$0.0: F15, K16; G29.6$+$0.1: K16; G31.9$+$0.0: \citealt{frail96,reachrho99,reach02}, H08, F15, K16; G32.4$+$0.1: K16; G32.8$-$0.1: \citealt{koralesky98,zhouchen11}; F15; G33.2$-$0.6: F15, K16; G33.6$+$0.1: K16, \citealt{zhou16}; G34.7$-$0.4: \citealt{claussen97,seta98b,reach05}; H08;
G49.2$-$0.7: G97; \citealt{koomoon97ii}; H08.}
\tablenotetext{b}{ * Denotes a SNR with an assumed distance of 8.5 kpc (the International Astronomical Union recommended distance to the Galactic center)  because the source does not have good constraints on its distance.}
\tablenotetext{c}{Evidence of Explosion Type: N = Neutron star detection; E = Environment suggestive of core-collapse SNe (e.g., molecular cloud interaction, nearby H{\sc ii} regions); A = metal abundances from X-ray observations are consistent with core collapse SNe.}
\tablenotetext{d}{References:
(1) \citealt{reynolds06}; (2) \citealt{helfand03}; (3) \citealt{ranasingheleahy18}; (4) \citealt{tian07}; (5) \citealt{ranasinghe18}; (6) \citealt{zhouchen09};
(7) \citealt{leahytian08}; (8) \citealt{su14}; (9) \citealt{su15}; (10) \citealt{frail13}; (11) \citealt{vasisht97}; (12) \citealt{vasisht00}; 
(13) \citealt{ranasinghe17}; (14) \citealt{kilpatrick16}; (15) \citealt{yamaguchi04}; (16) \citealt{zhouchen11}; (17) \citealt{bamba16}; (18) \citealt{sato16}; (19) \citealt{auchettl14}; (20) \citealt{uchida12};  (21) \citealt{radhakrishnan72}; (22) \citealt{zhu13}; (23) \citealt{tianleahy13}; (24) \citealt{sasaki14}}
\vspace{-10mm}
\end{deluxetable*}

\begin{deluxetable*}{llcrcrrrc}[ht]
\footnotesize
\tablenum{2}
\tablecaption{List of SNRs in Our Sample from CGPS\label{tab:tableCGPS}}
\tablehead{
\colhead{No.} & \colhead{Source\tablenotemark{a}} & \colhead{Alternate Names} & \colhead{Distance\tablenotemark{b}} & \colhead{Evidence of} & \colhead{$P_{1}/P_{0}$} & \colhead{$P_{2}/P_{0}$} & \colhead{$P_{3}/P_{0}$} & \colhead{References\tablenotemark{d}} \\
\colhead{} & \colhead{} & \colhead{} & \colhead{(kpc)}
& \colhead{Explosion Type\tablenotemark{c}} & \colhead{($\times10^{-5}$)} & \colhead{($\times10^{-7}$)} & \colhead{($\times10^{-7}$)} & \colhead{}}
\startdata
23 & G65.1$+$0.6$^\dagger$ & -- & 9.2$^{+0.4}_{-0.2}$ & N, E & 50.1$_{-1.0}^{+0.9}$ & 1748$^{+19}_{-18}$ & 79.2$^{+2}_{-3}$ & 1 \\
24 & G67.7$+$1.8 & -- & 8.5* & A & 59.9$_{-1.2}^{+1.4}$ & 16.9$^{+2.7}_{-2.4}$ & 48.4$\pm2$ & 2 \\
25 & G69.0$+$2.7 & CTB~80 & 1.5$^{+0.6}_{-0.4}$ & N & 6040$\pm38$ & 20060$^{+181}_{-186}$ & 2073$^{+24}_{-22}$ & 3, 4 \\
26 & G69.7$+$1.0 & -- & 8.5* & -- & 11.7$\pm0.4$ & 67.6$\pm2.8$ & 4.16$^{+0.37}_{-0.40}$ & -- \\ 
27 & G73.9$+$0.9$^\dagger$ & -- & 1.3$^{+0.7}_{-0.8}$ & E & 22.6$\pm0.2$ & 25.4$^{+8.4}_{-8.3}$ & 12.4$\pm0.3$ & 5 \\ 
28 & G78.2$+$2.1 & $\gamma$~Cygni SNR & 2.0$^{+0.6}_{-0.3}$ & N & 77.1$\pm0.1$ & 1139$^{+1}_{-2}$ & 93.3$\pm0.2$ & 6, 7 \\
29 & G84.2$-$0.8 & -- & 6.0$\pm$0.2 & -- & 780$_{-8}^{+7}$ & 5151$^{+6}_{-5}$ & 87.3$^{+2.8}_{-3.0}$ & 8 \\ 
30 & G85.4$+$0.7 & -- & 3.5$\pm$1.0 & -- & 792$_{-8}^{+9}$ & 1429$\pm3$ & 88.9$^{+3.7}_{-4.2}$ & 9 \\ 
31 & G93.7$-$0.2 & CTB~104A & 1.5$\pm$0.2 & -- & 329$_{-1}^{+1}$ & 230$\pm3$ & 221$\pm2$ & 10 \\ 
32 & G94.0$+$1.0$^\dagger$ & 3C~434.1 & 4.5$\pm$1.5 & E & 126$\pm1$ & 489$\pm2$ & 5.08$\pm0.14$ & 11, 12 \\ 
33 & G106.3$+$2.7 & -- & 0.8$^{+1.2}_{-0.1}$ & N, E & 50.0$_{-0.6}^{+0.7}$ & 6010$^{+4}_{-3}$ & 104$\pm2$ & 13 \\ 
34 & G109.1$-$1.0$^{\dagger}$ & CTB~109 & 3.2$\pm$0.2 & N, E & 30.7$\pm0.1$ & 739$^{+3}_{-2}$ & 12.1$^{+0.2}_{-0.1}$ & 14 \\ 
35 & G114.3$+$0.3 & -- & 0.7$^{+0.9}_{-0.0}$ & N & 53.1$\pm0.2$ & 128$\pm1$ & 1.66$^{+0.07}_{-0.06}$ & 15, 16 \\ 
36 & G116.5$+$1.1 & -- & 1.6$\pm$0.6 & -- & 24.3$\pm0.3$ & 1458$\pm9$ & 12.6$^{+0.5}_{-0.4}$ & 15 \\ 
37 & G116.9$+$0.2 & CTB~1 & 1.6$^{+1.5}_{-0.0}$ & A &  87.7$\pm0.4$ & 331$^{+3}_{-2}$ & 10.6$\pm0.3$ & 15, 17 \\ 
38 & G120.1$+$1.4$^{\diamond}$ & Tycho & 2.4$^{+2.6}_{-0.9}$ & A, L & 8.00$\pm0.03$ & 0.09$\pm0.01$ & 0.65$\pm0.01$ & 18, 19 \\
39 & G127.1$+$0.5$^{\dagger}$ & R~5 & 1.2$\pm$0.1 & E &  20.6$\pm0.1$ & 124$\pm1$ & 10.3$\pm0.2$ & 20 \\ 
40 & G132.7$+$1.3$^{\dagger}$ & HB~3 & 2.2$\pm$0.2 & E, A &  308$_{-0.4}^{+0.5}$ & 583$\pm2$ & 74.0$\pm0.4$ & 17, 21 \\ 
41 & G160.9$+$2.6 & HB~9 & 0.8$^{+1.0}_{-0.4}$ & -- &  47.7$\pm0.1$ & 364$\pm1$ & 0.33$\pm0.02$ & 22 \\ 
42 & G166.0$+$4.3$^{\dagger}$ & VRO~42.05.01 & 4.5$\pm$1.5 & E &  1.13$_{-0.10}^{+0.09}$ & 745$^{+11}_{-9}$ & 63.8$^{+1.4}_{-1.4}$ & 23 
\enddata
\tablenotetext{a}{ $\diamond$ Denotes SNRs thought to be from Type Ia SNe. $\dagger$ Denotes SNRs with evidence of interaction with a molecular cloud: G65.1$+$0.6: F15; G73.9$+$0.9: \citealt{zdz16}; G94.0$+$1.0: \citealt{jeong13}; G109.1$-$1.0: \citealt{sasaki06}; G127.1$+$0.5: \citealt{zhou14}; G132.7$+$1.3: K16, \citealt{zhou16b}}
\tablenotetext{b}{ * Denotes a SNR with an assumed distance of 8.5 kpc (the International Astronomical Union recommended distance to the Galactic center)  because the source does not have good constraints on its distance.}
\tablenotetext{c}{Evidence of Explosion Type: N = Neutron star detection; E = Environment suggestive of core-collapse SNe (e.g., molecular cloud interaction, nearby H{\sc ii} regions); A = metal abundances from X-ray observations; L = light echo spectrum.}
\tablenotetext{d}{References: (1) \citealt{tian06}; (2) \citealt{hui09}; (3) \citealt{li05}; (4) \citealt{leahy12}; (5) \citealt{loz93}; (6) \citealt{leahy13}; (7) \citealt{hui15}; (8) \citealt{leahy12b}; (9) \citealt{jackson08}; (10) \citealt{uya02}; (11) \citealt{foster05}; (12) \citealt{jeong13}; (13) \citealt{kothes01}; (14) \citealt{kothes12}; (15) \citealt{yar04}; (16) \citealt{kulkarni93}; (17) \citealt{lazendic06}; (18) \citealt{tian11}; (19) \citealt{krause08}; (20) \citealt{leahy06}; (21) \citealt{routledge91}; (22) \citealt{leahy07}; (23) \citealt{landecker89}}
\end{deluxetable*}

\begin{deluxetable*}{llcrcrrrc}[ht]
\footnotesize
\tablenum{3}
\tablecaption{List of SNRs in Our Sample from MOST\label{tab:tableMOST}}
\tablehead{
\colhead{No.} & \colhead{Source\tablenotemark{a}} & \colhead{Alternate Names} & \colhead{Distance\tablenotemark{b}} & \colhead{Evidence of} & \colhead{$P_{1}/P_{0}$} & \colhead{$P_{2}/P_{0}$} & \colhead{$P_{3}/P_{0}$} & \colhead{References\tablenotemark{d}} \\
\colhead{} & \colhead{} & \colhead{} & \colhead{(kpc)}
& \colhead{Explosion Type\tablenotemark{c}} & \colhead{($\times10^{-5}$)} & \colhead{($\times10^{-7}$)} & \colhead{($\times10^{-7}$)} & \colhead{}}
\startdata
43 & G289.7$-$0.3 & -- & 8.5* & -- & 71.8$^{+0.4}_{-0.3}$ & 971$\pm5$ & 0.04$\pm0.01$ & \\
44 & G290.1$-$0.8$^{\dagger}$ & MSH~11$-$61A & 7.0$\pm$1.0 & N, E, A & 23.8$\pm0.1$ & 230$\pm1$ & 9.40$\pm0.07$ & 1, 2, 3 \\ 
45 & G294.1$-$0.0 & -- & 8.5* & -- & 788$^{+11}_{-10}$ & 6189$^{+134}_{-121}$ & 2376$^{+37}_{-42}$ & \\ 
46 & G296.1$-$0.5 & -- & 3.0$\pm$1.0 & A & 523$\pm2$ & 2578$^{+16}_{-19}$ & 688$^{+5}_{-6}$ & 4, 5 \\ 
47 & G296.8$-$0.3 & -- & 9.6$\pm$0.6 & -- & 1134$\pm3$ & 3630$\pm17$ & 373$\pm3$ & 6 \\ 
48 & G298.6$-$0.0$^{\dagger}$ & -- & 8.5* & E & 57.4$\pm0.1$ & 1285$^{+4}_{-3}$ & 121$\pm1$ & 7 \\
49 & G299.6$-$0.5 & -- & 8.5* & -- & 115$\pm1$ & 67.3$^{+3.0}_{-2.6}$ & 77.7$^{+1.3}_{-1.5}$ & \\
50 & G301.4$-$1.0 & -- & 8.5* & -- & 79.6$^{+1.8}_{-2.0}$ & 2422$\pm1$ & 591$^{+14}_{-13}$ & \\
51 & G302.3$+$0.7$^{\dagger}$ & -- & 8.5* & E & 48.2$^{+0.7}_{-0.6}$ & 946$\pm10$ & 567$^{+3}_{-4}$ & 8 \\
52 & G304.6$+$0.1$^{\dagger}$ & Kes~17 & 9.7$^{+4.3}_{-1.7}$ & E, A &  30.1$\pm0.1$ & 137$\pm1$ & 6.71$\pm0.03$ & 9, 10 \\ 
53 & G308.1$-$0.7 & -- & 8.5* & -- & 12.5$\pm0.4$ & 74.8$^{+3.0}_{-3.4}$ & 141$\pm2$ & \\
54 & G308.8$-$0.1 & -- & 6.9$^{+8.1}_{-2.9}$ & N & 447$\pm1$ & 2612$^{+7}_{-6}$ & 221$\pm1$ & 11 \\
55 & G309.2$-$0.6 & -- & 4.0$\pm$2.0 & -- & 4.93$\pm0.09$ & 2278$^{+6}_{-5}$ & 65.5$^{+3.3}_{-3.1}$ & 12 \\
56 & G309.8$+$0.0 & -- & 8.5* & -- & 503$\pm2$ & 1237$^{+17}_{-16}$ & 1418$^{+7}_{-8}$ & -- \\
57 & G310.6$-$0.3 & Kes~20B & 8.5* & -- & 228$\pm1$ & 1005$\pm6$ & 47.8$\pm0.6$ & -- \\ 
58 & G310.8$-$0.4 & Kes~20A & 13.7 & E & 209$\pm1$ & 4222$^{+9}_{-7}$ & 509$\pm2$ & 13, 14 \\
59 & G311.5$-$0.3$^{\dagger}$ & -- & 14.8 & E & 1.92$\pm0.02$ & 9.75$^{+0.14}_{-0.16}$ & 0.28$\pm0.02$ & 14 \\ 
60 & G312.4$-$0.4$^{\dagger}$ & -- & 6.0$^{+8.0}_{-0.0}$ & E & 441$\pm1$ & 172$^{+4}_{-3}$ & 151$\pm1$ & 15 \\ 
61 & G315.4$-$2.3$^{\diamond}$ & RCW~86 & 2.5$^{+0.3}_{-0.2}$ & E, A & 116$\pm6$ & 4379$\pm2$ & 2098$^{+7}_{-6}$ & 16, 17, 18  \\ 
62 & G316.3$-$0.0 & MSH~14$-$57 & 7.2$\pm$0.6 & -- & 2.28$^{+0.05}_{-0.04}$ & 228$^{+1}_{-2}$ & 61.1$\pm0.3$ & 10 \\
63 & G317.3$-$0.2 & -- & 8.5* & -- & 267$\pm1$ & 7315$\pm15$ & 781$\pm3$ & \\
64 & G318.2$+$0.1 & -- & 8.5* & -- & 66.6$^{+0.9}_{-1.0}$ & 2537$^{+21}_{-23}$ & 120$^{+4}_{-3}$ & \\
65 & G321.9$-$0.3 & -- & 6.5$^{+3.5}_{-1.0}$ & N & 289$\pm2$ & 1588$^{+18}_{-16}$ & 85.9$^{+1.9}_{-2.2}$ & 19 \\ 
66 & G321.9$-$1.1 & -- & 8.5* & -- & 2051$^{+25}_{-23}$ & 7226$^{+122}_{-155}$ & 3639$\pm6$ & \\
67 & G322.5$-$0.1 & -- & 8.5* & N & 50.0$^{+0.5}_{-0.6}$ & 257$^{+5}_{-4}$ & 112$\pm1$ & 20 \\ 
68 & G323.5$+$0.1 & -- & 8.5* & -- & 21.4$^{+0.1}_{-0.2}$ & 460$\pm3$ & 51.6$^{+0.5}_{-0.4}$ & -- \\
69 & G326.3$-$1.8 & MSH~15$-$56 & 4.1$\pm$0.7 & N & 35.6$\pm0.1$ & 114$\pm1$ & 0.60$\pm0.02$ & 16, 21 \\ 
70 & G327.1$-$1.1 & -- & 8.5$\pm$0.5 & N & 45.3$\pm0.3$ & 126$^{+1}_{-2}$ & 11.8$\pm0.2$ & 22 \\ 
71 & G327.4$+$0.4 & Kes~27 & 4.3$^{+1.1}_{-0.0}$ & N & 237$\pm1$ & 1194$^{+5}_{-4}$ & 113$\pm1$ & 23, 24 \\ 
72 & G327.4$+$1.0 & -- & 8.5* & -- & 79.2$\pm0.4$ & 1040$^{+5}_{-4}$ & 240$\pm1$ & -- \\
\enddata
\tablenotetext{a}{ $\diamond$ Denotes SNRs thought to be from Type Ia SNe. $\dagger$ Denotes SNRs with evidence of interaction with a molecular cloud: G290.1$-$0.8: \citealt{filipovic05}; G298.6$-$0.0: \citealt{acero16}; G302.3$+$0.7: \citealt{frail96} (F96); G304.6$+$0.1: F96; \citealt{hewitt09}; G311.5$-$0.3: \citealt{andersen11}; G312.4$-$0.4: F96; G332.4$+$0.1: F96; G332.4$-$0.4: F96; \citealt{paron06}; G337.0$-$0.1: F96; G337.8$-$0.1: \citealt{koralesky98,zhang15}; G346.6$-$0.2: \citealt{koralesky98,hewitt09,andersen11}; G348.5$+$0.1: F96, \citealt{reynoso00}}
\tablenotetext{b}{ * Denotes a SNR with an assumed distance of 8.5 kpc (the International Astronomical Union recommended distance to the Galactic center)  because the source does not have good constraints on its distance.}
\tablenotetext{c}{Evidence of Explosion Type: N = Neutron star detection; E = Environment suggestive of core-collapse SNe (e.g., molecular cloud interaction, nearby H{\sc ii} regions); A = metal abundances from X-ray observations; L = light echo spectrum.}
\tablenotetext{d}{References: (1) \citealt{reynoso06}; (2) \citealt{auchettl15}; (3) \citealt{kaspi97}; (4) \citealt{castro11}; (5) \citealt{longmore77}; (6) \citealt{gaensler98b}; (7) \citealt{bamba16b}; (8) \citealt{frail96}; (9) \citealt{washino16}; (10) \citealt{caswell75}; (11) \citealt{caswell92}; (12) \citealt{rakowski01}; (13) \citealt{reach06}; (14) \citealt{andersen11}; (15) \citealt{doherty03}; (16) \citealt{rosado96}; (17) \citealt{sollerman03}; (18) \citealt{williams11}; (19) \citealt{stewart93}; (20) \citealt{whiteoak96}; (21) \citealt{temim13}; (22) \citealt{sun99}; (23) \citealt{mcclure01}; (24) \citealt{chen08}; (25) \citealt{park09}; (26) \citealt{vink04}; (27) \citealt{reynoso04}; (28) \citealt{frank15}; (29) \citealt{kaspi96}; (30) \citealt{eger11}; (31) \citealt{rakowski06}; (32) \citealt{yamaguchi14}; (33) \citealt{kothes07}; (34) \citealt{giacani11}; (35) \citealt{yamaguchi12}; (36) \citealt{koralesky98}; (37) \citealt{tian12}; (38) \citealt{halpern10}; (39) \citealt{tian07b}; (40) \citealt{giacani09}}
\end{deluxetable*}

\begin{deluxetable*}{llcrcrrrc}[ht!]
\footnotesize
\tablenum{3}
\tablecaption{List of SNRs in Our Sample from MOST (continued) \label{tab:tableMOST2}}
\tablehead{
\colhead{No.} & \colhead{Source\tablenotemark{a}} & \colhead{Alternate Names} & \colhead{Distance\tablenotemark{b}} & \colhead{Evidence of} & \colhead{$P_{1}/P_{0}$} & \colhead{$P_{2}/P_{0}$} & \colhead{$P_{3}/P_{0}$} & \colhead{References\tablenotemark{d}} \\
\colhead{} & \colhead{} & \colhead{} & \colhead{(kpc)}
& \colhead{Explosion Type\tablenotemark{c}} & \colhead{($\times10^{-5}$)} & \colhead{($\times10^{-7}$)} & \colhead{($\times10^{-7}$)} & \colhead{}}
\startdata 
73 & G329.7$+$0.4 & -- & 8.5* & E & 106$\pm1$ & 5.63$^{+0.05}_{-0.04}$ & 64.6$^{+0.8}_{-0.7}$ & -- \\
74 & G330.2$+$1.0 & -- & 4.9$^{+5.0}_{-0.0}$ & N & 144$\pm1$ & 114$\pm1$ & 24.1$^{+0.2}_{-0.3}$ & 23, 25 \\ 
75 & G332.0$+$0.2 & -- & 8.5* & -- & 0.17$\pm0.01$ & 770$\pm2$ & 60.6$\pm0.3$ & -- \\
76 & G332.4$+$0.1$^{\dagger}$ & Kes~32 & 7.5$^{+3.5}_{-0.9}$ & E & 18.2$\pm0.1$ & 1658$\pm2$ & 142$\pm1$ & 26 \\ 
77 & G332.4$-$0.4$^{\dagger}$ & RCW~103 & 3.3$^{+1.3}_{-0.2}$ & N, E, A & 20.2$^{+2.8}_{-3.2}$ & 12.8$\pm0.1$ & 4.29$\pm0.02$ & 27, 28 \\
78 & G335.2$+$0.1 & -- & 1.8 & N & 19.8$^{+10.2}_{-12.0}$ & 43.7$\pm0.6$ & 91.1$\pm0.4$ & 29, 30 \\ 
79 & G336.7$+$0.5 & -- & 8.5* & -- & 534$^{+62}_{-54}$ & 2092$\pm6$ & 241$\pm1$ & -- \\
80 & G337.2$-$0.7$^{\diamond}$ & -- & 2.0$\pm$0.5 & A & 4.43$\pm0.02$ & 3.71$^{+0.07}_{-0.08}$ & 2.75$\pm0.03$ & 31, 32 \\ 
81 & G337.3$+$1.0 & Kes~40 & 8.5* & -- & 2.15$\pm0.02$ & 165$\pm1$ & 34.7$^{+0.2}_{-0.1}$ & -- \\
82 & G337.8$-$0.1$^{\dagger}$ & Kes~41 & 11.0 & E & 5.51$\pm0.02$ & 1017$\pm1$ & 20.4$^{+0.1}_{-0.1}$ & 33 \\ 
83 & G340.4$+$0.4 & -- & 8.5* & -- & 93.0$\pm0.2$ & 188$\pm1$ & 68.4 $\pm0.4$ & -- \\ 
84 & G340.6$+$0.3 & -- & 15.0 & -- & 39.1$\pm0.2$ & 126$\pm1$ & 11.6$\pm0.1$ & 33 \\
85 & G341.9$-$0.3 & -- & 8.5* & -- & 204$\pm1$ & 385$\pm3$ & 24.3$^{+0.3}_{-0.4}$ & -- \\ 
86 & G342.0$-$0.2 & -- & 8.5* & -- & 50.0$\pm0.2$ & 203$\pm2$ & 9.25$^{+0.17}_{-0.21}$ & -- \\
87 & G342.1$+$0.9 & -- & 8.5* & -- & 13.0$\pm0.2$ & 218$^{+3}_{-2}$ & 2.98$\pm0.16$ & -- \\ 
88 & G343.1$-$0.7 & -- & 8.5* & -- & 34.4$^{+0.2}_{-0.3}$ & 2504$\pm9$ & 516$\pm3$ & -- \\ 
89 & G344.7$-$0.1$^{\diamond}$ & -- & 6.3$^{+7.7}_{-0.1}$ & A & 52.9$\pm0.2$ & 63.7$^{+0.6}_{-0.7}$ & 14.3$^{+0.1}_{-0.2}$ & 34, 35 \\ 
90 & G346.6$-$0.2$^{\dagger}$ & -- & 11 & E & 4.48$\pm0.04$ & 26.1$\pm0.3$ & 5.38$^{+0.06}_{-0.07}$ & 36 \\ 
91 & G348.7$+$0.3 & CTB~37B & 13.2$\pm$0.2 & N & 578$\pm1$ & 88.3$^{+3.0}_{-2.7}$ & 43.5$^{+0.3}_{-0.4}$ & 37, 38 \\ 
92 & G351.2$+$0.1 & -- & 8.5* & -- & 41.4$\pm0.1$ & 333$\pm1$ & 19.7$\pm0.1$ & \\
93 & G351.7$+$0.8 & -- & 13.2$\pm$0.5 & -- & 131$\pm1$ & 261$\pm2$ & 2.88$\pm0.14$ & 39 \\ 
94 & G351.9$-$0.9 & -- & 8.5* & -- & 270$\pm1$ & 533$\pm5$ & 92.2$\pm1.1$ & \\
95 & G352.7$-$0.1$^{\diamond}$ & -- & 7.5$^{+0.9}_{-0.7}$ & A & 0.07$\pm0.01$ & 218$\pm1$ & 36.1$\pm0.1$ & 40, 41 \\ 
96 & G354.8$-$0.8 & -- & 8.5* & -- & 11.7$\pm0.3$ & 3750$\pm2$ & 520$\pm3$ & \\
\enddata
\tablenotetext{a}{ $\diamond$ Denotes SNRs thought to be from Type Ia SNe. $\dagger$ Denotes SNRs with evidence of interaction with a molecular cloud: G290.1$-$0.8: \citealt{filipovic05}; G298.6$-$0.0: \citealt{acero16}; G302.3$+$0.7: \citealt{frail96} (F96); G304.6$+$0.1: F96; \citealt{hewitt09}; G311.5$-$0.3: \citealt{andersen11}; G312.4$-$0.4: F96; G332.4$+$0.1: F96; G332.4$-$0.4: F96; \citealt{paron06}; G337.0$-$0.1: F96; G337.8$-$0.1: \citealt{koralesky98,zhang15}; G346.6$-$0.2: \citealt{koralesky98,hewitt09,andersen11}; G348.5$+$0.1: F96, \citealt{reynoso00}}
\tablenotetext{b}{ * Denotes a SNR with an assumed distance of 8.5 kpc (the International Astronomical Union recommended distance to the Galactic center)  because the source does not have good constraints on its distance.}
\tablenotetext{c}{Evidence of Explosion Type: N = Neutron star detection; E = Environment suggestive of core-collapse SNe (e.g., molecular cloud interaction, nearby H{\sc ii} regions); A = metal abundances from X-ray observations; L = light echo spectrum.}
\tablenotetext{d}{References: (1) \citealt{reynoso06}; (2) \citealt{auchettl15}; (3) \citealt{kaspi97}; (4) \citealt{castro11}; (5) \citealt{longmore77}; (6) \citealt{gaensler98b}; (7) \citealt{bamba16b}; (8) \citealt{frail96}; (9) \citealt{washino16}; (10) \citealt{caswell75}; (11) \citealt{caswell92}; (12) \citealt{rakowski01}; (13) \citealt{reach06}; (14) \citealt{andersen11}; (15) \citealt{doherty03}; (16) \citealt{rosado96}; (17) \citealt{sollerman03}; (18) \citealt{williams11}; (19) \citealt{stewart93}; (20) \citealt{whiteoak96}; (21) \citealt{temim13}; (22) \citealt{sun99}; (23) \citealt{mcclure01}; (24) \citealt{chen08}; (25) \citealt{park09}; (26) \citealt{vink04}; (27) \citealt{reynoso04}; (28) \citealt{frank15}; (29) \citealt{kaspi96}; (30) \citealt{eger11}; (31) \citealt{rakowski06}; (32) \citealt{yamaguchi14}; (33) \citealt{kothes07}; (34) \citealt{giacani11}; (35) \citealt{yamaguchi12}; (36) \citealt{koralesky98}; (37) \citealt{tian12}; (38) \citealt{halpern10}; (39) \citealt{tian07b}; (40) \citealt{giacani09}; (41) \citealt{sezer14}}
\end{deluxetable*}

Barrel-shaped or bilateral SNRs, a subgroup of shell-type SNRs, are characterized by an axisymmetric morphology with two bright limbs. \cite{gaensler98} analyzed a sample of bilateral SNRs at radio frequencies and showed that their axes tended to be aligned with the Galactic plane. More recently, \cite{west16} investigated all Milky Way bilateral SNRs, and they showed that a simple model of SNRs expanding into an ambient Galactic magnetic field could reproduce the observed radio morphologies. 

To quantify the complex and varied morphologies of SNRs, \citealt{lopez09bkgnd} developed and applied several mathematical tools. Using the power-ratio method (PRM), \cite{lopez09,lopez11} showed that the thermal X-ray emission of Type Ia SNRs is more symmetric and circular than that of core-collapse SNRs. Subsequently, \cite{peters13} extended this approach to infrared images of SNRs and found similar results as in the X-ray. Recently, \cite{tyler17} used the PRM to compare the SNR soft X-ray morphologies to neutron star velocities and showed that the neutron stars are moving opposite to the bulk of the SN ejecta in many sources. For a detailed summary of these results and those of other groups, see \cite{lopez18}. 

In this paper, we investigate the radio morphologies of SNRs in the Milky Way to examine how asymmetries evolve with size and age. SNRs are observable for \hbox{$\sim10^{4}-10^{5}$}~years at radio wavelengths \citep{sarbadhicary17}, and their morphologies are shaped by interactions with the surrounding medium (e.g., \citealt{zhang18}) and by the magnetic field (e.g., \citealt{orlando07,west17}).

This paper is structured as follows. In Section~\ref{sec:data}, we describe the radio data and introduce our sample of Galactic SNRs. Section~\ref{sec:methods} outlines the power-ratio method which we employ to measure the asymmetries of the sources. Finally, Section~\ref{sec:results} presents our results and discusses the implications regarding SNR evolution.

\section{Data and Sample} \label{sec:data}

Our sample is comprised of 96 SNRs imaged in three surveys/catalogs of the Galactic plane: the HI, OH, Recombination Line (THOR) Survey \citep{Beuther16}, the Canadian Galactic Plane Survey (CGPS; \citealt{kothes06}), and the Molonglo Observatory Synthesis Telescope (MOST) SNR catalog \citep{whiteoak96} We describe the data and the sample selection from each survey below.

The final sample includes 59 SNRs with constraints on their explosive origins (5 Type Ia SNRs, 54 core-collapse SNRs). 63 out of the 96 SNRs have distance measurements in the literature, as listed in Tables~\ref{tab:tableTHOR}--\ref{tab:tableMOST}. Distances are determined primarily through H{\sc i} absorption spectra (e.g. \citealt{leahy14}) and/or via kinematic velocities of associated molecular clouds (e.g., \citealt{ranasinghe17,ranasinghe18}). The error bars on the distances in Tables~\ref{tab:tableTHOR}--\ref{tab:tableMOST} reflect the uncertainties quoted in the literature for each measurement. If no error bars are given for the distances in Tables~\ref{tab:tableTHOR}--\ref{tab:tableMOST}, then the references did not assess the uncertainties in those values.

\begin{figure*}
\centering
\includegraphics[width=\textwidth]{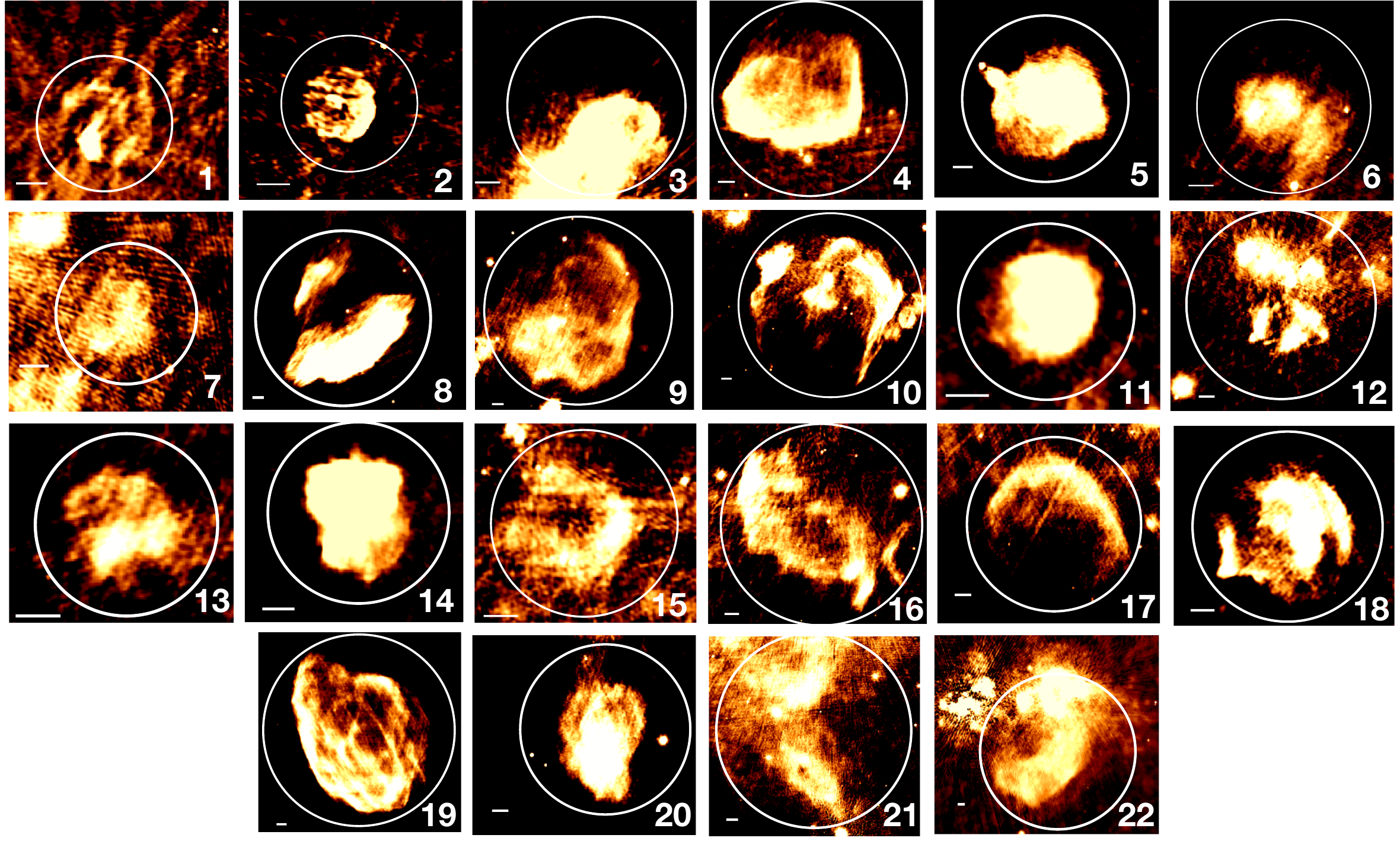}
\caption{\label{fig:thorSNRs} Radio continuum (1.4~GHz) images of 22 SNRs from the HI, OH, Recombination Line (THOR) Survey with the JVLA \citep{Beuther16}. The SNRs range in radius of 1.5--20\arcmin\ \citep{green17}. The extents of the sources are marked with white circles, and the white scale bar represents 2\arcmin. The numbers correspond to those in Column 1 of Table~\ref{tab:tableTHOR}.}
\end{figure*}

\subsection{THOR Survey}

The THOR Survey used the Karl G. Jansky Very Large Array \footnote{http://www.mpia.de/thor} to observe the radio continuum in 6 bands (from 1--2~GHz), the H{\sc i} 21-cm line, four OH lines, and radio recombination lines over the first Galactic Quadrant (Galactic longitudes from 14.5$^{\circ} < l <  67.25^{\circ}$ and latitudes of $\mid b \mid \leq 1.25^{\circ}$\footnote{Only the first half of the survey data is publicly available currently, which covers longitudes from 14.5 to 37.9 degrees and 47.1 to 51.2 degrees.}). We opted to analyze the 1.4~GHz radio data, since 95\% of SNRs are detected at this frequency \citep{chomiuk09,dubner15}. The spatial resolution of the THOR data is 20\arcsec.

In the area of the THOR survey, 34 SNRs have been identified \citep{green17}, and all were fully imaged. However, we excluded 12 SNRs due to substantial artifacts in the data (e.g., G15.4$+$0.1, G31.5$-$0.6) or because their radio emission is dominated by a pulsar wind nebula rather than the synchrotron from their shells (e.g., G21.5$-$0.9, Kes~75). The THOR sample of 22 SNRs is listed in Table~\ref{tab:tableTHOR}, and the images of the SNRs are presented in Figure~\ref{fig:thorSNRs}. The THOR SNRs have a range in radii from 1.5--20\arcmin\ \citep{green17}. 

19 out of the 22 THOR SNRs are likely from core-collapse explosions, based on the presence of neutron stars, metal abundances, and/or their dense, star-forming environments (see Table~\ref{tab:tableTHOR}). The other three have insufficient data to characterize explosion type (G21.5$-$0.1, G33.2$-$0.6, and G36.6$-$0.7), but given their location within the Galactic plane, they may also be from core-collapse supernovae (SNe). As many Type Ia SNRs are found at high galactic latitudes (e.g., Kepler, SN~1006), it is not surprising that the THOR sample has no known Type Ia sources. We note that two of the THOR SNRs considered in this work (Kes~69 and G28.6$-$0.1) were in the bilateral sample analyzed by \cite{west16}.

\subsection{CGPS}

\begin{figure*}[ht]
\centering
\includegraphics[width=0.75\textwidth]{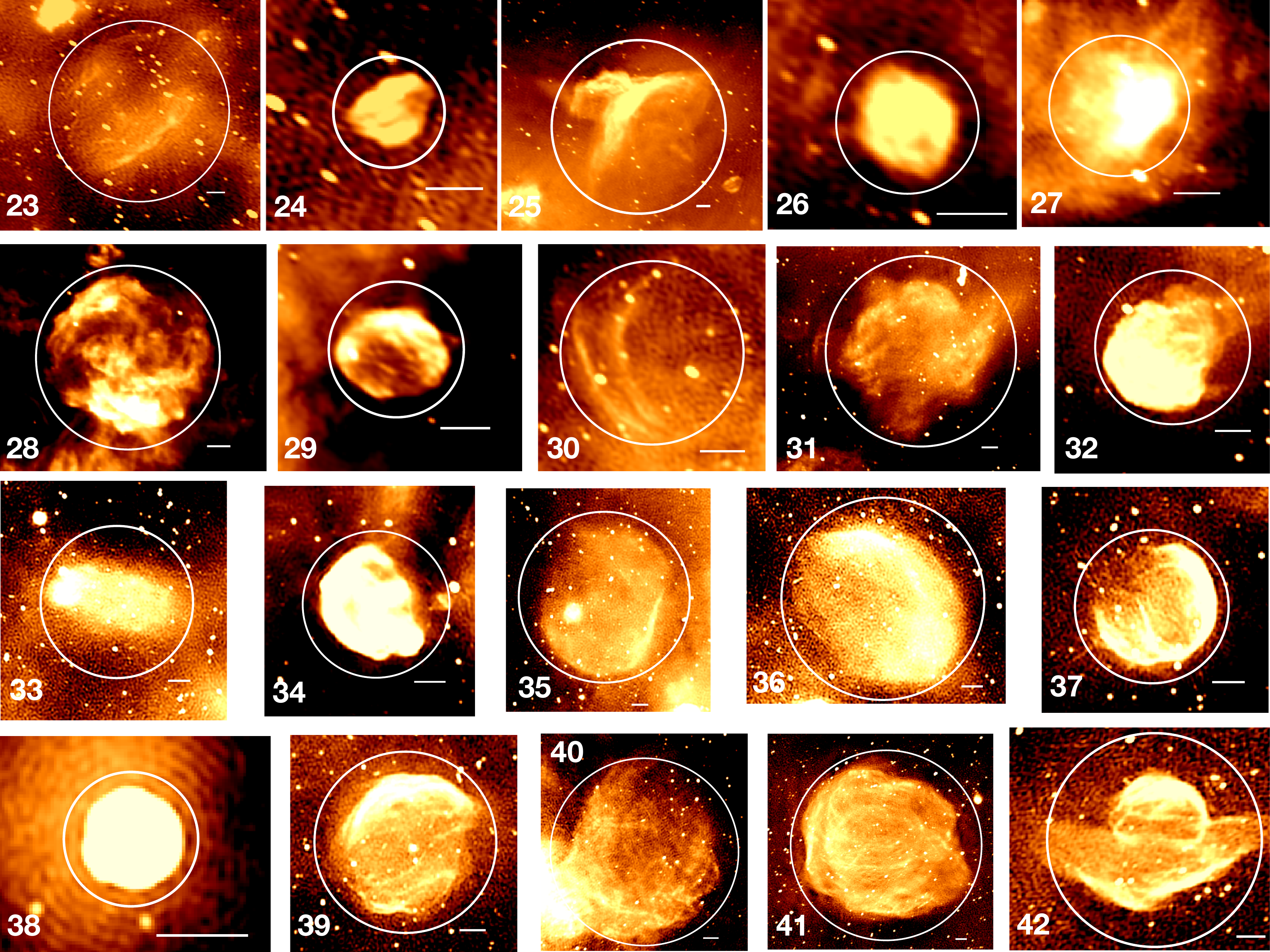}
\caption{\label{fig:cgpsSNRs} Radio continuum (1.42~GHz) images of 20 SNRs from the CGPS \citep{kothes06} in our sample. The SNRs have radii of 7--76\arcmin. The extents of the sources are marked with white circles, and the white scale bar represents 2\arcmin. The numbers correspond to those in Column 1 of Table~\ref{tab:tableCGPS}.}
\end{figure*}

The CGPS surveyed atomic hydrogen and the radio continuum using the Dominion Radio Astrophysical Observatory \citep{taylor03}. CGPS data is available in 5$^{\circ} \times 5^{\circ}$ mosaics\footnote{http://www.cadc-ccda.hia-iha.nrc-cnrc.gc.ca/en/cgps/} and covers longitudes of 52$^{\circ} < l <  192^{\circ}$ and latitudes of $-3.5^{\circ} < b < 5.5^{\circ}$. \cite{kothes06} catalogued the known and candidate Galactic SNRs in the CGPS fields, with a final sample of 36 sources. For our work, we analyzed the 1.42~GHz radio continuum CGPS data, which has 18\arcsec\ pixels and 1\arcmin\ resolution.

We downloaded the CGPS data and examined the 36 SNRs identified by \cite{kothes06}. We excluded 16 SNRs from the sample because of bright PWNe dominating their emission (e.g., G65.7$+$1.2 [DA~495], G130.7$+$3.1 [3C~58]), contaminating emission from foreground or nearby sources (e.g., G83.0$-$0.3), or low signal (e.g., G126.2$+$1.6). In addition, one source (G84.9$+$0.5) is now classified as a H{\sc ii} region \citep{foster07}, so it is not considered. The CGPS sample of 20 SNRs is listed in Table~\ref{tab:tableCGPS}, and the images of the SNRs are presented in Figure~\ref{fig:cgpsSNRs}. The CGPS SNRs span a range in radii of 7--76\arcmin.

\begin{figure*}
\centering
\includegraphics[width=0.8\textwidth]{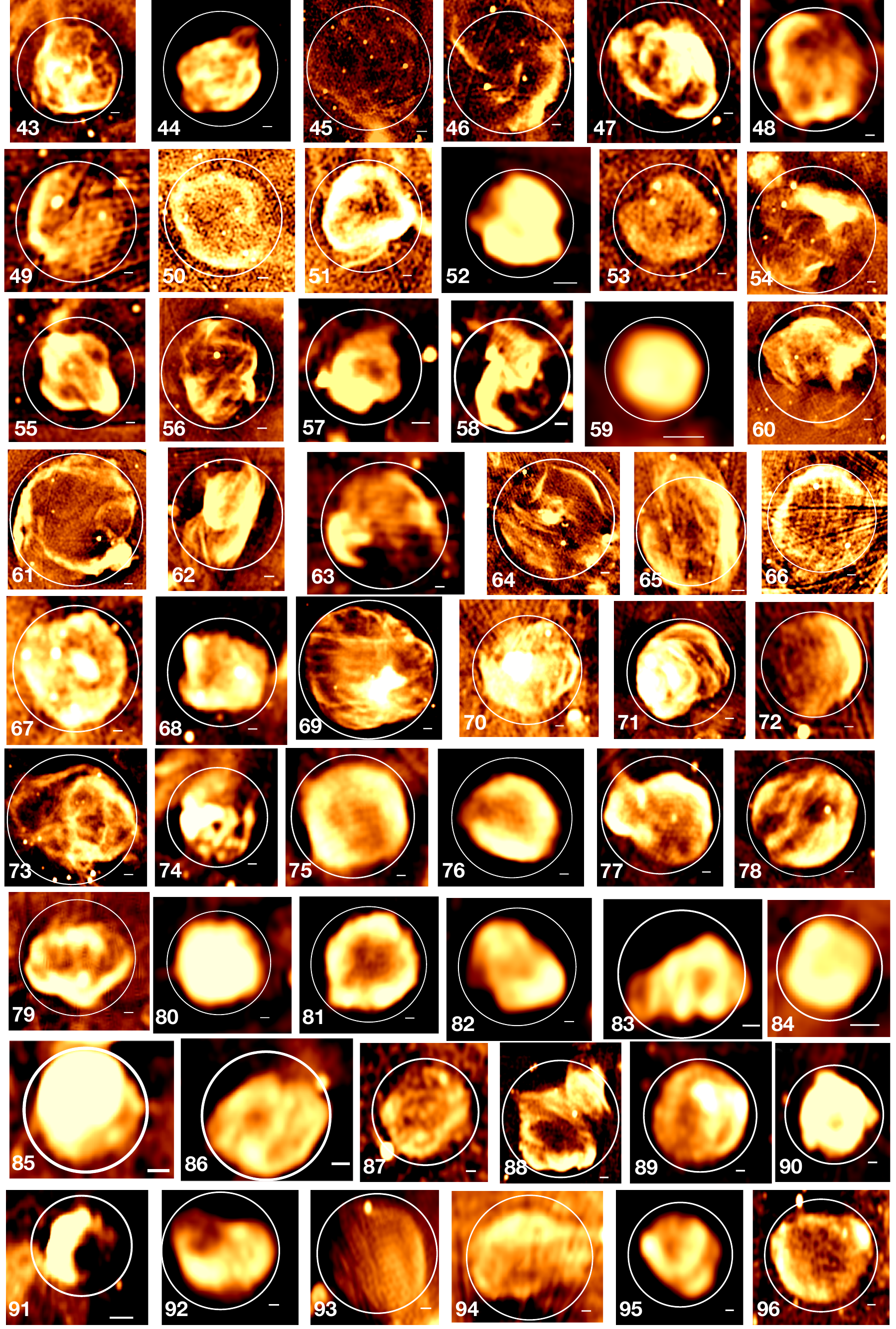}
\caption{\label{fig:mostSNRs} Radio continuum (0.843~GHz) images of 54 SNRs from the MSC \citep{whiteoak96} in our sample. The SNRs have radii of 3--23\arcmin. The extents of the sources are marked with white circles, and the white scale bar represents 2\arcmin. The numbers correspond to those in Column 1 of Table~\ref{tab:tableMOST}.}
\end{figure*}

13 of the 20 SNRs are classified as core-collapse explosions, based on detections of neutron stars, metal abundances, and/or interactions with dense, star-forming environments. One SNR (G120.1$+$1.4 [Tycho]) is associated with SN~1572 and known to have originated from a Type Ia explosion based on its light echo spectrum \citep{krause08}. The other six SNRs have insufficient data to characterize their explosion types. We note that four of the CGPS SNRs considered in this work (G65.1$+$0.6, G116.9$+$0.2 [CTB~1], G127.1$+$0.5, G166.0$+$4.3 [VRO~42.05.01]) were among the bilateral sample of \cite{west16}.

\subsection{MOST}

The MOST Galactic plane survey observed the radio continuum at 0.843~GHz and covers longitudes of 245$^{\circ} < l <  355^{\circ}$ and latitudes of $\mid b \mid \leq 1.5^{\circ}$. Each field of the survey covers an area of 70\arcmin $\times$ 70\arcmin\ and achieved spatial resolutions of 43\arcsec. \cite{whiteoak96} constructed the MOST SNR catalog (MSC) which included 57 known SNRs (the category called MSC.A in \citealt{whiteoak96}) and 18 new SNRs (the category called MSC.B in \citealt{whiteoak96}).

We downloaded the MOST data and examined the 75 SNRs from MSC.A and MSC.B. As with the THOR and CGPS samples, we excluded the MSC SNRs with emission dominated by PWNe (e.g., G291.0$-$0.1, G320.4$-$1.2, G338.3$-$0.0), with contamination from nearby sources (e.g., G315.4$-$2.3, G320.6$-$1.6, G349.2$-$0.1), or that were not sufficiently detected or resolved (e.g., G298.5$-$0.3, G345.7$-$0.2, G349.7$+$0.2). The final sample of 54 MSC SNRs analyzed in this work are listed in Table~\ref{tab:tableMOST}, and their images are given in Figure~\ref{fig:mostSNRs}. The MSC SNRs have radii of 3--23\arcmin.

Compared to the THOR and CGPS observed SNRs, the MSC SNRs in our sample have fewer constraints on their explosive origins, with 28 SNRs lacking observational indications of SN type. 22 MSC SNRs show evidence of being from core-collapse explosions, and four MSC SNRs are thought to be from Type Ia SNe (RCW~86, G337.2$-$0.7, G344.7$-$0.1, G352.7$-$0.1; see references in Table~\ref{tab:tableMOST}). Seven of the MSC SNRs in our sample (G302.3$+$0.7, G317.3$-$0.2, G321.9$-$0.3, G327.4$+$1.0, G332.0$+$0.2, G332.4$-$0.4, G354.8$-$0.8) overlap with the bilateral sample of \cite{west16}.

\subsection{MAGPIS}

We also considered a sample of 60 SNRs using 1.4~GHz data from the Multi-Array Galactic Plane Imaging Survey (MAGPIS) \citep{helfand06}. However, we found that the noise in these data led to large uncertainties in the derived power ratios. Thus, we did not this sample in this paper.

\section{Methods} \label{sec:methods}

We measure the symmetry of our sample using a multipole expansion technique called the power-ratio method (PRM). This technique was developed to quantify the morphologies of galaxy clusters \citep{buote95,buote96,jeltema05}. Subsequently, it was extended to measure the asymmetries of SNRs in X-ray and infrared images \citep{lopez09,lopez11,peters13,tyler17}. Our work here applies the same technique to radio observations of SNRs.  An overview of the method is provided below. For a more detailed description including the mathematical formalism, we refer readers to \cite{lopez09,lopez11}.

The PRM measures asymmetries by calculating the multipole moments of emission in a circular aperture. It is derived in a similar way to the expansion of a two-dimensional gravitational potential, except an image's surface brightness replaces the mass surface density. The powers $P_{m}$ are obtained by integrating the magnitude of each term of the multiple expansion over a circle of radius $R$. We divide the powers $P_{m}$ by $P_{0}$ to normalize with respect to flux, and we set the origin position in our apertures to the geometric centers of the SNR's radio emission. In this case, each term of the multipole expansion reflects asymmetries at successively smaller scales. The dipole power ratio $P_{1}/P_{0}$ represents the bulk asymmetry of the SNR's emission. The quadrupole power ratio $P_{2}/P_{0}$ measures the ellipticity or elongation of a source, and the octupole power ratio $P_{3}/P_{0}$ quantifies the mirror asymmetry.

Before applying the PRM, we removed the point sources from the images by replacing the sources' pixel values with those interpolated from surrounding background regions. We determined the SNRs' geometric centers by putting a circle at their right ascensions and declinations reported by \cite{green17} and adjusting those positions and sizes to encompass their radio extents. For each SNR, we then ran the PRM on the source and on the adjacent, source-free background and subtracted the background moments from the SNRs' moments.

The selection of the geometric center influences the derived power ratios. For example, we computed the power ratios for G49.2$-$0.7 (W51C) after shifting the centroid 10 and 20 pixels (25\arcsec\ and 50\arcsec, respectively) from the original analysis to explore the associated uncertainties. The largest effect was on $P_{1}/P_{0}$,  which had a 10--60\% (4--260\%) change for the 10-pixel (20-pixel) shift. By comparison, the fractional changes in $P_{2}/P_{0}$ and $P_{3}/P_{0}$ were 2--16\% (5--38\%) and 12--28\% (23--30\%), respectively. These values depend on the direction and magnitude of the offset as well as the sub-structure and morphology of the individual sources.

The physical resolution (parsecs per pixel) -- which depends on the assumed distance to the targets -- affects the derived power ratios as well. To demonstrate this point, we calculated the power ratios for G49.2$-$0.7 (W51C) assuming distances of 0.5--1.5$\times$ the estimated distance of 5.4~kpc (i.e., adopted distances of 2.7--8.1~kpc). Shorter distances, corresponding to greater physical resolution, increases the derived power ratios: assuming a distance of 2.7~kpc, the power ratios increased by 43\%. At larger distances, the power ratios decrease: a distance of 8.1~kpc produced power ratios that are 22\% lower than those listed in Table~\ref{tab:tableTHOR}. Thus, the derived power ratios are somewhat sensitive to the adopted distances to the individual targets.

To estimate the uncertainties in the power ratios from noise in the images, we used the Monte Carlo approach as described in \cite{lopez09bkgnd}. Specifically, we smoothed out noise using the program {\it AdaptiveBin} \citep{sanders01}, and then we added noise back into the images assuming that each pixel intensity is mean of a Poisson distribution, and selecting an intensity from that distribution. We repeated this procedure 100 times to create 100 mock images of each source. We ran the PRM on the 100 images, and we calculated the power ratios from the mean of these 100 values. The 1-$\sigma$ confidence limits are derived from the sixteenth highest and lowest PR values.

\begin{figure*}
\includegraphics[width=\textwidth]{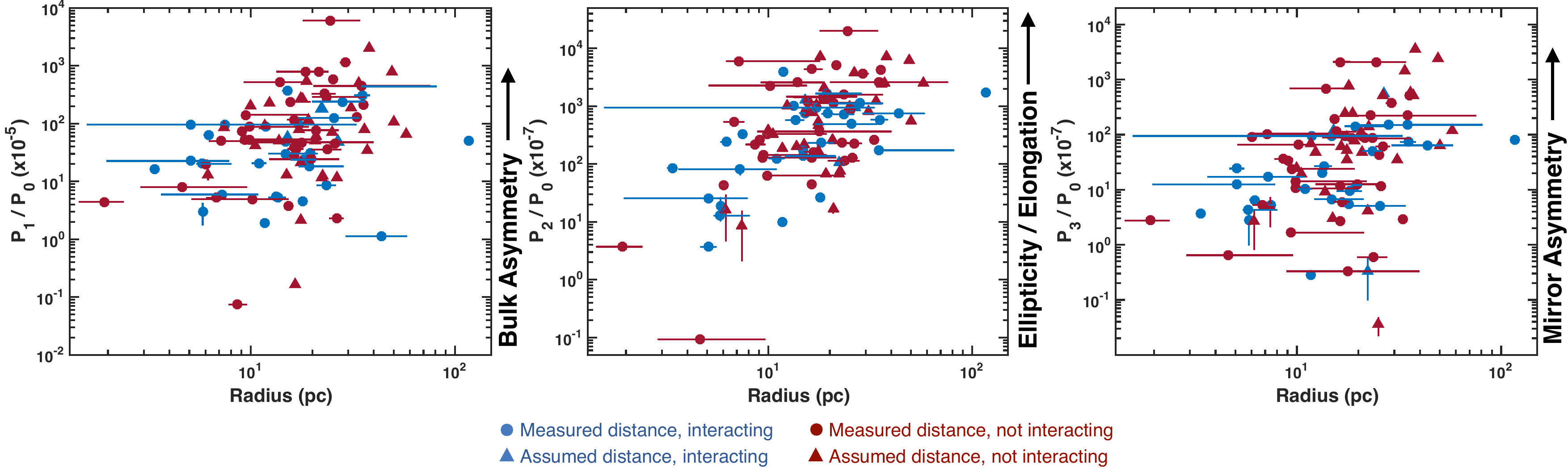} 
\caption{Dipole power ratio $P_{1}/P_{0}$ (left), quadrupole power ratio $P_{2}/P_{0}$ (middle), and octupole power ratio $P_{3}/P_{0}$ (right) versus radius. $P_{1}/P_{0}$ measures bulk asymmetry, $P_{2}/P_{0}$ quantifies ellipticity/elongation, and $P_{3}/P_{0}$ reflects mirror asymmetries. As shown in the legend, red and blue colors represent whether SNRs have no evidence or some evidence of interaction with molecular clouds, respectively. Symbol shapes denote whether each source has a measured distance in the literature (circles) or no known distance (triangles). Horizontal error bars represent the uncertainty in radii given the sources' distance estimates.}
\label{fig:P2}
\end{figure*}

\section{Results and Discussion} \label{sec:results}

In Figure \ref{fig:P2}, we plot the dipole power ratio ($P_{1}/P_{0}$; left), the quadrupole power ratio ($P_{2}/P_{0}$; middle), and the octupole power ratio ($P_{3}/P_{0}$; right) versus radius. To convert the radii to parsecs, we adopt the distances listed in Tables~\ref{tab:tableTHOR}--\ref{tab:tableMOST} and the angular extents given in \cite{green17}. 

\begin{deluxetable}{ccc}
\tablenum{4}\tablecolumns{3}
\tablecaption{Median Power-Ratio Results by Radius\label{tab:results}}
\tablehead{\colhead{Power-Ratio} & \colhead{Radius $\lesssim$10~pc} & \colhead{Radius $\gtrsim$10~pc}}
\startdata
\cutinhead{All 96 SNRs}
$P_{1}/P_{0}$ & $2.0\times10^{-4}$ & $6.7\times10^{-4}$ \\
$P_{2}/P_{0}$ & $8.1\times10^{-6}$ & $7.5\times10^{-5}$ \\
$P_{3}/P_{0}$ & $5.3\times10^{-7}$ & $6.8\times10^{-6}$ \\
\cutinhead{63 SNRs with Measured Distances}
$P_{1}/P_{0}$ & $2.0\times10^{-4}$ & $5.1\times10^{-4}$ \\
$P_{2}/P_{0}$ & $8.5\times10^{-6}$ & $7.4\times10^{-5}$ \\
$P_{3}/P_{0}$ & $6.5\times10^{-7}$ & $6.5\times10^{-6}$ \\
\enddata
\end{deluxetable}

The sample spans a wide range of power-ratio values, and the median power ratios of the SNRs by radius are listed in Table ~\ref{tab:results}. Generally, SNRs with radii $\lesssim$10~pc have smaller power ratios than those with radii $\gtrsim$10~pc, and this trend is most pronounced in $P_{2}/P_{0}$. We note that 33 of the 96 SNRs have no distance measurements to date (the triangles in Figure~\ref{fig:P2}), and we have assumed a distance of 8.5~kpc to those sources. However, if those SNRs are located at closer distances, their power ratios would increase (see Section~\ref{sec:methods}). Thus, we also list the median power ratios of only the 63 SNRs with measured distances in Table~\ref{tab:results}. We find that this sub-sample gives similar results, with the smaller SNRs giving lower power ratios, indicative of less asymmetries than the larger SNRs.

Comparing the SNRs associated with Type Ia versus those from core-collapse explosions, G337.2$-$0.7 and Tycho and  have among the lowest power-ratio values among both samples. However, the other Type Ia SNRs have near the median or greater power ratios than the CC SNRs. These results suggest that radio continuum morphology is not a reflection of explosion type, in contrast to X-ray and infrared morphologies where the two classes have distinct symmetries \citep{lopez09,lopez11,peters13}. Gven that G337.2$-$0.7 and Tycho have the smallest and third-smallest radii, respectively, of the 96 SNRs in the sample, their symmetric morphologies in the radio may simply reflect their young age and small size.

The radii $R_{\rm s}$ of the SNRs are a rough proxy of age $t$, since $R_{\rm s} \propto t^{m}$, where $m$ is the expansion parameter, and the shock velocity $v_{\rm s}$ is given by $v_{\rm s} = mR_{\rm s}/t$. The value of $m$ depends on the evolutionary stage of the SNR. During free expansion, $m \sim 1$, and as the shock begins to decelerate, $m \sim 0.6-0.8$ \citep{chevalier82a,chevalier82b}. Once the shock has swept-up a mass $M_{\rm sw}$ that is comparable to the mass of the ejecta $M_{\rm ej}$, then the SNR enters the Sedov-Taylor (ST) phase, when $m = 0.4$ \citep{sedov59}. Subsequently, $m$ decreases from $m = 0.33$ in the pressure-driven snowplow stage (e.g., \citealt{blondin98}) to $m = 0.25$ in the momentum-conserving stage \citep{cioffi88}. 

\begin{deluxetable}{lrrrrc}
\tablenum{5}
\tablecaption{Ages of the THOR SNRs\label{tab:ages}}
\tablehead{
\colhead{Source} & \colhead{$R_{\rm s}$\tablenotemark{a}} & \colhead{$n_{0}$} & \colhead{$M_{\rm SW}$} & \colhead{$t_{\rm kyr}$} & \colhead{References\tablenotemark{d}} \\
\colhead{} & \colhead{(pc)} & \colhead{(cm$^{-3}$)} & \colhead{($M_{\sun}$)} & \colhead{(kyr)} & \colhead{}}
\startdata
G15.9$+$0.2 & 7.4$\pm$3.5 & 0.7 & 41 & 2.9\tablenotemark{d} & 1 \\
G16.7$+$0.1 & 5.8$\pm$2.3 & 1.0\tablenotemark{b} & 28 & 1.5 & -- \\
G18.1$-$0.1 & 7.4$\pm$0.2 & 0.6 & 35 & 4.4\tablenotemark{d} & 2 \\
G18.8$+$0.3 & 28.1$\pm$8.1 & 1.0\tablenotemark{b} & 3211 & 76\tablenotemark{b} & -- \\
G20.0$-$0.2 & 16.3$\pm$0.4 & 1.0\tablenotemark{b} & 627 & 19\tablenotemark{b} & -- \\
G20.4$+$0.1 & 9.1$\pm$4.7 & 1.0\tablenotemark{b} & 109 & 4.5\tablenotemark{b} & -- \\
G21.5$-$0.1 & 6.2$\pm$2.9 & 1.0\tablenotemark{b} & 34 & 1.7\tablenotemark{b} & -- \\
G21.8$-$0.6 & 15.1$^{+0.0}_{-0.9}$ & 1.0\tablenotemark{b} & 498 & 16\tablenotemark{b} & -- \\
G22.7$-$0.2 & 16.6$\pm$1.5 & 1.0\tablenotemark{b} & 662 & 20\tablenotemark{b} & -- \\
G23.3$-$0.3 & 17.3$\pm$15.7 & 4.0 & 2997 & 45 & 3 \\
G27.4$+$0.0 & 3.4$\pm$0.2 & 0.6 & 3.4 & 1.1\tablenotemark{d} & 4 \\
G28.6$-$0.1 & 15.4$\pm$0.5 & 0.2 & 106 & 7.5\tablenotemark{d} & 5 \\
G29.6$+$0.1 & 7.3$\pm$3.7 & 1.0\tablenotemark{b} & 40 & 2.6\tablenotemark{c} & -- \\
G31.9$+$0.0 & 6.2$\pm$0.3 & 2.0 & 69 & 3.5\tablenotemark{d} & 6 \\
G32.4$+$0.1 & 14.8$\pm$3.5 & 1.0\tablenotemark{b} & 469 & 15 & -- \\
G32.8$-$0.1 & 11.9$\pm$9.9 & 0.1 & 24 & 4.2\tablenotemark{d} & 7 \\
G33.2$-$0.6 & 22.3$\pm$10.5 & 1.0\tablenotemark{b} & 1605 & 43 & -- \\
G33.6$+$0.1 & 5.1$\pm$0.4 & 0.4 & 7.7 & 3.0\tablenotemark{d} & 8 \\
G34.7$-$0.4 & 13.5$\pm$1.4 & 5.0 & 1780 & 10 & 9 \\
G35.6$-$0.4 & 6.8$\pm$0.8 & 1.0\tablenotemark{b} & 46 & 2.2 & -- \\
G36.6$-$0.7 & 30.9$\pm$14.5 & 1.0\tablenotemark{b} & 4269 & 96 & -- \\
G49.2$-$0.7 & 23.6$\pm$2.6 & 0.1 & 190 & 18\tablenotemark{d} & 10
\enddata
\tablenotetext{a}{The error bars on $R_{\rm s}$ reflect the uncertainties in the source distances from Table~\ref{tab:tableTHOR}. If no uncertainties in distance are given in the literature (or if we have assumed a distance of 8.5~kpc), we assign a distance uncertainty of 4~kpc.}
\tablenotetext{b}{Assumed $n_{0} = 1.0$~cm$^{-3}$ as no constraints on density were found in the literature \citep{fer01}.}
\tablenotetext{c}{Assumed $E_{51} = 1.0$ as no constraints on explosion energy were found in the literature.}
\tablenotetext{d}{We have scaled the age estimates from the references to the distances listed in Table~\ref{tab:tableTHOR} and the SNR radii in this table.}
\tablenotetext{e}{References: (1) \citealt{reynolds06}; (2) \citealt{leahy14}; (3) \citealt{castro13}; (4) \citealt{kumar14}; (5) \citealt{bamba01}; (6) \citealt{chen04}; (7) \citealt{zhouchen11}; (8) \citealt{sun04}; (9) \citealt{reach05}; (10) \citealt{koo95}}
\end{deluxetable}

SNR ages are typically derived by assuming that SNRs are in the ST phase of their evolution, the stage that most radio-bright SNRs are thought to be observed \citep{berk86,berezhko04}. In this case, ages are determined based on the observed shock velocity by the expression $t = 2 R_{\rm s} / 5 v_{\rm s}$. Alternatively, given an estimate of the mass density of the ISM $\rho_{\rm o}$ and the explosion energy $E$, SNR ages can be derived using the ST solution \citep{sedov59}:

\begin{equation}
R_{\rm s}=1.15\bigg(\frac{E}{\rho_{\rm o}}\bigg)^{1/5}t^{2/5}=5.0~E_{51}^{1/5} n_{\rm o}^{-1/5} t_{\rm kyr}^{2/5}~\textrm{pc}.
\label{eq:sedov}
\end{equation}

\noindent
On the right-hand side of the equation, $E_{51}$ is the explosion energy in units of $10^{51}~{\rm erg}$ and $t_{\rm kyr}$ is the SNR age in kyr. 

To demonstrate the large uncertainties in age and the challenge of comparing those values to the power ratios, we compiled the age estimates of the THOR SNRs found in the literature (listed in Table~\ref{tab:ages}). When available (e.g., for G15.9$+$0.2, G32.8$-$0.1, G33.6$+$0.1), these values are the dynamical ages derived from the shock radius $R_{\rm s}$ and velocity $v_{\rm s}$, using the relation $t = 2 R_{\rm s} / 5 v_{\rm s}$. For those SNRs without constraints on $v_{\rm s}$, we scale the ages estimated from the ST solution in the references (see Table~\ref{tab:ages}) to the distances in Table~\ref{tab:tableTHOR} and radii in Table~\ref{tab:ages}. For those SNRs without any age estimates in the literature, we adopt the ISM densities $n_{\rm o}$ in Table~\ref{tab:ages} and assume explosion energies of $E_{51} = 1$ to estimate $t_{\rm kyr}$.

To evaluate the validity of the assumption that all of the SNRs are in the ST phase, we calculated the mass swept-up $M_{\rm SW}$ by their forward shocks: \hbox{$M_{\rm SW} = \frac{4}{3} \pi R_{\rm s}^3 \times 1.4 m_{\rm H} n_{\rm o}$}, where $m_{\rm H}$ is the mass of hydrogen. The resulting $M_{\rm SW}$ for each SNR is listed in Table~\ref{tab:ages}, along with the adopted shock radii $R_{\rm s}$ and ISM densities $n_{\rm o}$ to derive $M_{\rm SW}$. For two SNRs (G27.4$+$0.0 and G33.6$+$0.1), $M_{\rm SW} < 10 M_{\sun}$, and thus their forward shocks may have swept up less than their ejecta masses $M_{\rm ej}$. This result would indicate that they may not have reached the ST phase yet, so their age estimates $t_{\rm kyr}$ in Table~\ref{tab:ages} are upper limits. Five SNRs (G18.8$+$0.3, G23.3$-$0.3, G33.2$-$0.6, G34.7$-$0.4, G36.6$-$0.7) have $M_{\rm SW} > 10^{3} M_{\sun}$ and thus may have transitioned past the ST phase, which occurs at a time $t_{\rm tr} = 2.9\times10^{4} E_{51}^{4/17} n_{\rm o}^{-9/17}~{\rm years}$ when the shock has swept up $M_{\rm SW} \approx 10^{3} E_{51}^{15/17} n_{\rm o}^{-14/17}~M_{\sun}$ \citep{blondin98}. In these cases, the age estimates $t_{\rm age}$ should be interpreted as lower-limits.

\begin{figure*}
\includegraphics[width=\textwidth]{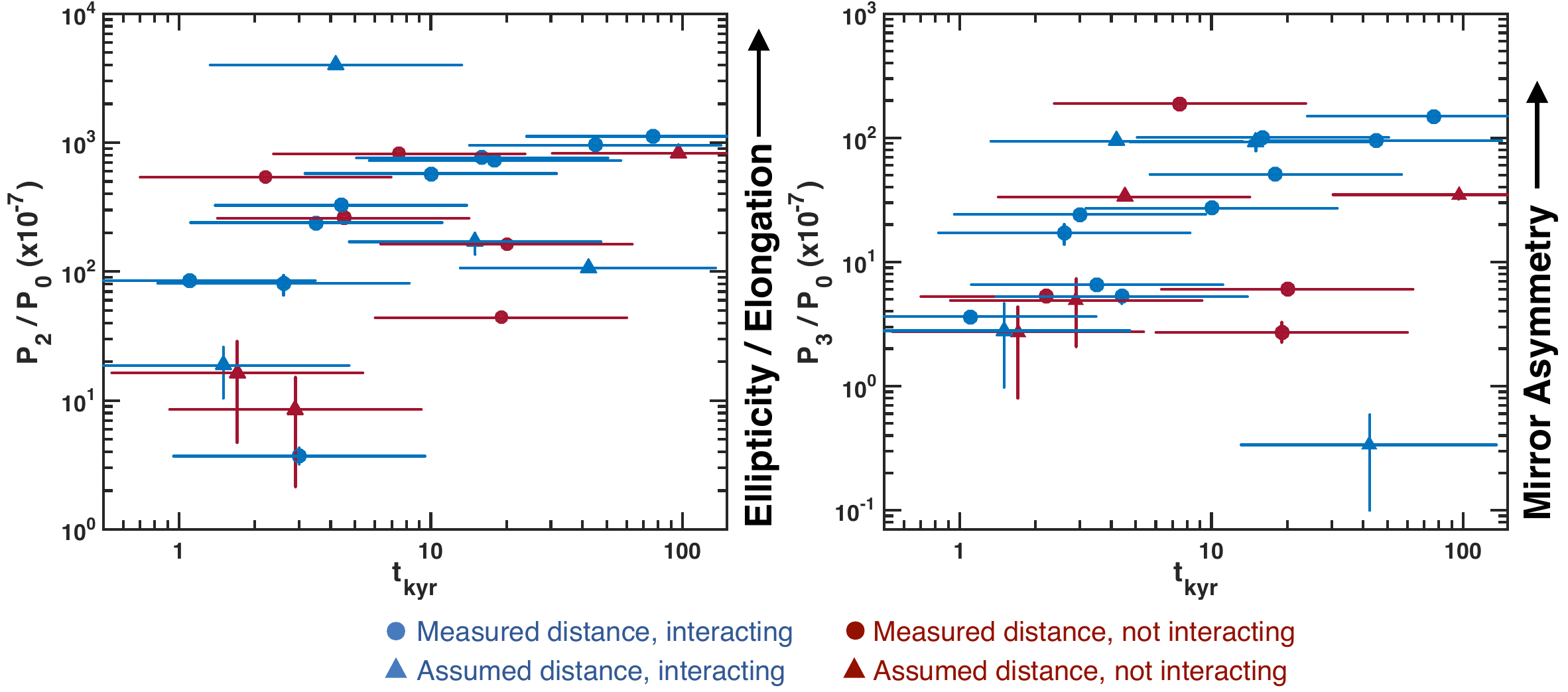}
\caption{Plots of $P_{2}/P_{0}$ (left) and $P_{3}/P_{0}$ (right) versus age $t_{\rm kyr}$ (listed in Table~\ref{tab:ages}), assuming the SNRs are in the Sedov-Taylor phase. Error bars on $t_{\rm kyr}$ are represent the uncertainty in the ambient density $n_{\rm o}$, which we conservatively assume to be one order of magnitude.}
\label{fig:age}
\end{figure*}

In Figure~\ref{fig:age}, we plot $P_{2}/P_{0}$ (left panel) and $P_{3}/P_{0}$ (right panel) versus age $t_{\rm kyr}$ of the THOR sample. We find a weak trend that younger SNRs ($\lesssim$3~kyr old) have lower $P_{2}/P_{0}$ and $P_{3}/P_{0}$ than the older SNRs ($\gtrsim$3~kyr old), consistent with the results shown in Figure~\ref{fig:P2}. These findings suggest that SNRs' forward shocks are initially more symmetric, and their expansion into an inhomogeneous medium increases the asymmetries with time. The large dispersion in the power-ratio values in the small/young SNRs indicates that the objects may begin with different degrees of asymmetry as well, possibly reflecting their explosion geometries or the inhomogeneous environments immediately surrounding the SN. 

One reason that the plots in Figure~\ref{fig:age} show less correlation compared to the power ratios versus radius in Figure~\ref{fig:P2} may be due to the large uncertainty in the SNR ages. Specifically, the explosion energies and the ISM densities are not well constrained in many cases. For many SNRs, we assumed $n_{\rm o} = 1$~cm$^{-3}$ (as an approximation of the density of the warm neutral medium: \citealt{fer01}) and $E_{51} = 1$ due to a lack of any observational constraints. Realistically, these parameters can range from $E_{51} \sim$ 0.1--10 (e.g., \citealt{sukhbold16}) and $n_{\rm o} \sim 10^{-3}$--10$^{2}$~cm$^{-3}$. In particular, $n_{\rm o}$ spans several orders of magnitude and depends on the environment of the SNR. If the SNR is expanding into a progenitor's wind bubble, then $n_{\rm o} \sim10^{-2}$--10$^{-3}$ (e.g., RCW~86: \citealt{broersen14}); if the SNR is interacting with a molecular cloud, then $n_{\rm o} \sim 10$--100~cm$^{-3}$ (e.g., gamma-ray bright SNRs: \citealt{castro10}). Generally, $n_{\rm o}$ can be estimated from modeling of X-ray or gamma-ray observations \citep{castro13}.

Thus, some age estimates could be off by a factor of $\sim$10 or more, shifting their placement in Figure~\ref{fig:age}. Given that the uncertainty in $n_{\rm o}$ is responsible for a large uncertainty in the ages $t_{\rm kyr}$, we compute the error bars of $t_{\rm kyr}$ by calculating the ages for an order-of-magnitude smaller and larger $n_{\rm o}$. Consequently, the horizontal error bars in Figure~\ref{fig:age} are conservative estimates.

Our results are consistent with recent three-dimensional, hydrodynamical simulations of SNRs expanding into an inhomogeneous medium \citep{kim15,martizzi15,walch15,zhang18}. In these works, the authors follow the evolution of SNRs in a multi-phase or turbulent ISM. They find that the SNRs in an inhomogeneous medium become progressively more asymmetric compared to those in a homogeneous ISM. For example, \cite{martizzi15} showed that the blast wave travels faster in the inhomogeneous medium case, particularly in areas of low-density channels around the SNR. \cite{zhang18} found that the mean ambient density is the primary factor influencing SNRs' evolution and that a smoother (lower Mach number) turbulent structure leads to faster, more asymmetric expansion\footnote{We note that \cite{martizzi15} and \cite{zhang18} had different results on the impact of turbulent structure on SNR expansion and morphology. \cite{martizzi15} showed that SNRs expand faster in a more turbulent medium, while \cite{zhang18} found that smoother turbulent structure leads to faster shock expansion. \cite{zhang18} attributes the disparity to how each study models turbulence: \cite{martizzi15} used a lognormal density distribution, whereas \cite{zhang18} adopted an initial Gaussian velocity perturbation in a uniform medium that grows with time.}. Thus, the increase in SNR asymmetries with radius (Figure~\ref{fig:P2}) and with age (Figure~\ref{fig:age}) reflect the inhomogeneous, turbulent structure of the ISM. 

We note that \cite{lopez09,lopez11} found no age evolution in the power-ratios derived from X-ray images of SNRs. However, the soft X-rays trace thermal emission from SNR ejecta, so SNR X-ray morphologies may reflect explosion asymmetries. Furthermore, these previous studies considered only young sources, with ages $t_{\rm kyr}\lesssim$10, whereas our sample spans a wider range, with $t_{\rm kyr}\sim$1--100. Further investigation is necessary to determine if size/age evolution is unique to the SNRs' radio continuum morphologies. 

To compute ages, we assumed a uniform ambient density, though our results indicate that the SNRs are expanding into inhomogeneous environments (since a homogeneous ISM would lead to no size/age evolution in the asymmetries). \cite{zhang18} showed that the mean ambient density is the dominant factor in determining shock expansion with time, though inhomogeneities are important in shaping SNRs overall. Thus, the assumption of a single $n_{\rm o}$ may be sufficient. However, given the large uncertainties in the SNR ages, the radii are the best observable indicator of SNR evolutionary stage. 

In this work, we have analyzed nearly one-third of the SNR population of the Milky Way. In the future, application and comparison to extragalactic SNRs may reveal differences in the turbulent structure of nearby galaxies. While the fractal nature of the Milky Way ISM \citep{elmegreen96} is also observed in nearby galaxies (e.g., in the Small Magellanic Cloud: \citealt{stan99}), the medium can differ substantially, e.g., in porosity \citep{bagetakos11} or in molecular gas velocity dispersion \cite{sun18}. Ultimately, comparison of SNRs' morphological evolution with simulations (e.g., \citealt{kim15,martizzi15,walch15,zhang18}) may place new constraints on the ISM properties of the Milky Way and nearby galaxies. 

\acknowledgements

We thank Drs. Carles Badenes and Adam Leroy for useful discussions. This work was supported through NSF Astronomy \& Astrophysics Grant AST$-$1517021. L.A.L. acknowledges support from the Sophie and Tycho Brahe Visiting Professorship at the Niels Bohr Institute. 

\bibliography{bibliography}

\end{document}